\documentclass[a4paper]{spie} 

\usepackage{amsmath} 
\usepackage{amssymb}	
\usepackage{graphicx} 
\usepackage[usenames,dvipsnames]{pstricks}
\usepackage{pst-grad} 
\usepackage{pst-plot} 
\usepackage{color}
\usepackage{wasysym}
\usepackage{url}
\usepackage{bbold}
\usepackage{bibentry}
\usepackage{bm}

\usepackage{bibentry}

\renewcommand{\v}[1]{\ensuremath{\mathbf{#1}}} 
\newcommand{\abs}[1]{\left| #1 \right|} 
\newcommand{\avg}[1]{\left< #1 \right>} 
\renewcommand{\d}[2]{\frac{d #1}{d #2}} 
\newcommand{\pdd}[2]{\frac{\partial^2 #1}{\partial #2^2}} 
\newcommand{\ket}[1]{\left| #1 \right>} 
\let\baraccent=\= 
\renewcommand{\=}[1]{\stackrel{#1}{=}} 



\title{Quantum confinement in Si and Ge nanostructures: Effect of crystallinity} 

\author{Eric G. Barbagiovanni\supit{a}, David J. Lockwood\supit{b}, Raimundo N. Costa Filho\supit{c}, Lyudmila V. Goncharova\supit{d}, Peter J. Simpson\supit{d}
\skiplinehalf
\supit{a}Laboratory for Simulation of Physical Systems, Beijing Computational Science Research Centre, Beijing 100084, People's Republic of China; \\
\supit{b}National Research Council, Ottawa, Ontario K1A 0R6, Canada; \\
\supit{c}Departamento de F\'{i}sica, Universidade Federal do Cear\'{a}, Caixa Postal 6030, Campus do Pici, 60455-760 Fortaleza, Cear\'{a}, Brazil; \\
\supit{d}Department of Physics and Astronomy, University of Western Ontario, London, Ontario N6A 3K7, Canada; \\
}

\authorinfo{E.G.B.: E-mail: santino@csrc.ac.cn; Affiliated with: Departamento de F\'{i}sica, Centro de F\'{i}sica das Interac\c{c}\~{o}es Fundamentais, Lisboa 1049-001, Portugal}
 
\begin{document} 
  \maketitle 

\begin{abstract}
We look at the relationship between the preparation method of Si and Ge nanostructures (NSs) and the structural, electronic, and optical properties in terms of quantum confinement (QC). QC in NSs causes a blue shift of the gap energy with decreasing NS dimension. Directly measuring the effect of QC is complicated by additional parameters, such as stress, interface and defect states. In addition, differences in NS preparation lead to differences in the relevant parameter set. A relatively simple model of QC, using a `particle-in-a-box'-type perturbation to the effective mass theory, was applied to Si and Ge quantum wells, wires and dots across a variety of preparation methods. The choice of the model was made in order to distinguish contributions that are solely due to the effects of QC, where the only varied experimental parameter was the crystallinity. It was found that the hole becomes de-localized in the case of amorphous materials, which leads to stronger confinement effects. The origin of this result was partly attributed to differences in the effective mass between the amorphous and crystalline NS as well as between the electron and hole. Corrections to our QC model take into account a position dependent effective mass. This term includes an inverse length scale dependent on the displacement from the origin. Thus, when the deBroglie wavelength or the Bohr radius of the carriers is on the order of the dimension of the NS the carriers `feel' the confinement potential altering their effective mass. Furthermore, it was found that certain interface states (Si-O-Si) act to pin the hole state, thus reducing the oscillator strength.
\end{abstract}


\section{INTRODUCTION}\label{intro}

Semiconductor nanostructures (NSs) exhibit increased oscillator strength due to electron hole wave function overlap, and band gap engineering due to the effect of quantum confinement (QC). Thus, materials like Si are a viable option for opto-electronics, photonics, and quantum computing.\cite{edit:2010, Lockwood:2004, Loss:1998} QC is defined as the modification in the free particle dispersion relation as a function of a system's spatial dimension.\cite{Yoffe:2002} If a free electron is confined within a potential barrier, a shift in the band gap energy is observed, which is inversely proportional to the system size squared, in the effective mass approximation. As a result, the emitted photon energy is directly proportional to the gap energy ($E_G$). QC often manifests itself in optical experiments when the dimension of the system is systematically reduced and an increase in the absorbed/emitted photon energy is measured corresponding to electron transitional states.

For practical applications, utilizing QC effects in NSs requires an understanding of the band structure of a low-dimensional material, how the method of preparation effects the final properties of the NS, and the kinetics/ dynamics of the absorption/emission process. The confinement potential is determined by the alignment of the respective Fermi levels when a material of a $E_{G1}$ is surrounded by a material of a $E_{G2}$, with $E_{G1}<E_{G2}$.\cite{Frensley:1977} The preparation technique can introduce stress in the system, which changes the band gap energy.\cite{Bir:1974} For indirect gap materials phonon processes can effect the recombination mechanism.\cite{Valentin:2007} The lifetime associated with the recombination event can be altered by the excitation power.\cite{VanDao:2005} (For a review of general properties of low-dimensional structures, see Refs. \citenum{Heiss:2005, Lockwood:2004, Yoffe:2002}. For a discussion of other higher order effects in NSs, see Ref. \citenum{Heiss:2005}.) This article is concerned with the electron/hole recombination process in amorphous (a) versus crystalline (c) NSs with different dimensions.

Several theoretical models (e.g. see Refs. \citenum{Zunger:2001, Tran:1990, Tit:2010}) have been applied to NS; all models are empirical and no one model can model all semiconductor NSs. Since the parameters of a NS system are dependent upon the preparation method for a particular material, a comprehensive theoretical understanding must test along this dimension as well. In this article, we consider a relatively simple model of direct e-h recombination using a `particle in a box' type model as a perturbation to the effective mass theory. We use no adjustable parameters and include corrections to the model dependent on the preparation method as known experimentally and/or computationally when needed, thus achieving transparency in the physics involved. The only parameter tested in this work is the crystallinity, which is shown to effect the strength of confinement (defined in Sec. \ref{theory}), because of the different symmetry properties of the electron and hole. 

The model is applied to experimental results on crystalline and amorphous Si and Ge NSs, including quantum wells (QWs), wires (Q-wires) and dots (QDs). Systems of regular shape are chosen to ensure crystallinity is the primary parameter.  For example, data obtained by van Buuren et al. \cite{Buuren:1998} for high quality `star-shaped' samples are difficult to analyse theoretically. Parameters relevant to a particular system are discussed and energy corrections are given when needed. Results are discussed and a mechanism for the differences between the strength of confinement in the amorphous and crystalline system is proposed. Central to the discussion of the confinement strength in NSs is how the effective mass (EM) changes as a function of dimension. Recently, Cosentino et al.\cite{Cosentino:2013} demonstrated an increase in the confinement strength due to a reduced EM. We developed a model for the EM as a function of dimension and considered its effect in a QW. 

Modifications to the EM parameter include the use of a spatially dependent effective mass (SPDEM) introduced via the von Roos Hamiltonian \cite{vonRoos:1983}. This form of the EM is appropriate for doped semiconductors, or when there exists a graded potential \cite{vonRoos:1983, Young:1989, Geller:1993}. At the interface of a NS where the crystal potential will change, possibly abruptly, the EM will change. The influence of the interface on the EM is expressed through the Bastard type boundary conditions (B.Cs) \cite{Bastard:1981, Chetounai:1995, Ganguly:2006, Borovitskaya:2000, Moskalenko:2007}. Proper treatment of the EM in a NS is an unresolved problem. The EM is important for theoretical models \cite{Tomic:2011, Delerue:2004, Niquet:2000_1}, and related to the hopping parameter and carrier mobility in the tight-binding model \cite{Seino:2012}. There is experimental \cite{Barbagiovanni:2012, Seas:1997, Cosentino:2013} and theoretical \cite{Seino:2011} evidence that the EM should depend on NS dimension. Fundamentally, since QC causes an increase in $\v{k}$-space \cite{Barbagiovanni:2012}, one does expect an increase in the bulk EM ($m_o^{*-1}(\propto\pdd{E}{\v{k}})$). Furthermore, a change in the $m_o^*$ will modify the Bohr radius, therefore, altering the regime in which QC effects can be observed. The challenge is that experimental measurements of the $m_o^*$ in a NS are model dependent \cite{Lockwood:1996, Rossner:2003}, and it is difficult to theoretically scale the EM in a NS \cite{Barbagiovanni:2012, Seino:2012}. However, the EM provides a natural framework to incorporate the influence of a modified crystal potential due to the interface, which is not adequately accounted for in theoretical models \cite{Delerue:2004, Barbagiovanni:2013, Seino:2011}.

Recently, Costa Filho et al. addressed the problem of a SPDEM by introducing a characteristic inverse length scale into the translation operator \cite{CostaFilho:2011}, given by:
\begin{equation}\label{eq1}
\mathcal{T}_{\gamma}(a)\ket{x}=\ket{x+a+\gamma a x}.
\end{equation}
$\gamma$ defines the inverse length scale and mixes the displacement, $a$, of a carrier particle with the original position, $x$. Note that, this formalism is equivalent to the $q$-exponential formalism in Tsallis nonextensive thermostatistics \cite{CostaFilho:2011} and results in a deformed or contracted space. Tsallis thermostatistics describes a meta-equilibrium system, which provides a natural connection to confined carrier particles in an excited state. From Eq. \eqref{eq1} a modified momentum operator was derived:
\begin{equation}\label{eq2}
\hat{p}_{\gamma}=-i\hbar(1+\gamma x)\d{}{x};
\end{equation}
along with a definition for the SPDEM \cite{CostaFilho:2011}:
\begin{equation}\label{eq3}
m(x)=\frac{m_o^*}{\left(1+\gamma x\right)^2}.
\end{equation}
Furthermore, it was shown that a point canonical transformation (PCT) of a Hamiltonian defined with $\hat{p}_{\gamma}$ in the kinetic term and with a harmonic potential energy yields an effective Hamiltonian for a particle with constant mass in a Morse potential \cite{CostaFilho:2013}, in perfect analogy with Ref. \citenum{vonRoos:1983}. Therefore, Eq. \eqref{eq3} provides a natural framework to describe the anharmonic influence of the potential barrier on the carrier particles in a NS. In this report, we consider a SPDEM given by Eq. \eqref{eq3} and discuss the physical implications of this term on the confinement energy for a QW. Note that the discussion, presented here generalizes to the case of a quantum dot and wire.

\section{THEORY}\label{theory}

The Bohr radius of an electron (e), hole (h) or exciton (X) is given by, in SI units:
$$
a_{\text{e}(\text{h})(X)}=\frac{4\pi\epsilon\hbar^2}{m^*_{\text{e}(\text{h})(X)}e^2},
$$
$m^*_{\text{e}(\text{h})(X)}$ is the effective mass of the e, h or X, respectively, $e$ is the electric charge and $\epsilon$ is the dielectric constant. Depending on the e or h effective mass, the X-Bohr radius is 4.5 nm for Si and 24 nm for Ge. The Bohr radius defines the spatial dimension of the particles, which determines the range of sizes for which QC can be observed. We define three regimes of confinement here:\cite{Yoffe:2002}
\begin{itemize}
\item Weak confinement: When the dimension of the system is much larger than $a_{\text{e}}$ and $a_{\text{h}}$. In this situation, the appropriate mass in the kinetic term is $M=m^*_{\text{e}}+m^*_{\text{h}}$. The energy term is dominated by the Coulomb energy. 
\item Medium confinement: When the dimension of the system is much smaller than $a_{\text{e}}$, but larger than $a_{\text{h}}$, then only electrons will experience confinement. The relevant mass is simply $m_{\text{e}}^*$ for the kinetic term. Most materials belong to this class.
\item Strong confinement: When the dimension of the system is much smaller than $a_{\text{e}}$ and $a_{\text{h}}$. Here both electrons and holes experience confinement and the relevant mass is the reduced mass, $\mu$, with $\frac{1}{\mu}=\frac{1}{m^*_{\text{e}}}+\frac{1}{m^*_{\text{h}}}$. In this regime, the Coulomb term is small and can generally be treated as a perturbation. 
\end{itemize}
Below we will use the terms `weak,' `medium' and `strong' to refer to the different regimes of confinement discussed above.

In the `particle-in-a-box' model the bulk $E_G$ is taken as the ground state energy. The effect of reduced dimension is considered as a perturbation to the bulk $E_G$ within the EM approximation. We consider an infinite confinement model and use the bulk EM (the details of the model are in Refs. \citenum{Barbagiovanni:2011, Barbagiovanni:2012}). The expression for the variation of $E_G$ is given by:
\begin{equation}\label{eq4}
E_{G}(D)=E_{G}(\infty)+\frac{A}{D^2}\,\text{eV}\cdot\text{nm}^2.\\
\end{equation}
$E_{G}(\infty)$ is the band gap of the bulk material and $D$ represents the QD diameter, the QW thickness or the Q-Wire diameter in what follows. The calculation was carried out for confinement in 1D, 2D with cylindrical coordinates and 3D with spherical coordinates. The parameter $A$ is given for Si and Ge in the strong, medium and weak confinement regimes in Table \ref{tbl1}. The change in energy of the conduction band minimum ($\Delta$E$_{CBM}$) due to QC is labelled as `medium confinement' in Table \ref{tbl1}, because a $\Delta$E$_{CBM}$ is equivalent to QC of the electron only as defined by our model, where only the electron mass is considered in Eq. \eqref{eq4}. The change in energy of the valence band maximum ($\Delta$E$_{VBM}$) due to QC is also listed in Table \ref{tbl1}, which is calculated by considering confinement of the hole only, where only the hole mass is considered in Eq. \eqref{eq4}. The other fixed parameter is the appropriate $E_G(\infty)$ of the bulk system.
\begin{table}
\caption{Parameter $A$  given in Eq. \eqref{eq4} for 3D, 2D, 1D confinement and for $\Delta$E$_{CBM}$, $\Delta$E$_{VBM}$.}
\label{tbl1}
\begin{center}
\begin{tabular}{c c c c} 
{}& {} & {Si} & {Ge}\\
\hline
{3D}&{Strong}&{3.57}&{7.88}\\
{}&{Medium ($\Delta$E$_{CBM}$)}&{1.39}&{2.69}\\
{}&{Weak}&{0.91}&{1.77}\\
{}&{$\Delta$E$_{VBM}$}&{-2.64}&{-5.19}\\
\hline
{2D}&{Strong}&{2.09}&{4.62}\\
{}&{Medium ($\Delta$E$_{CBM}$)}&{0.81}&{1.58}\\
{}&{Weak}&{0.53}&{1.04}\\
{}&{$\Delta$E$_{VBM}$}&{-1.55}&{-3.04}\\
\hline
{1D}&{Strong}&{0.89}&{1.97}\\
{}&{Medium ($\Delta$E$_{CBM}$)}&{0.35}&{0.67}\\
{}&{Weak}&{0.23}&{0.44}\\
{}&{$\Delta$E$_{VBM}$}&{-0.66}&{-1.30}
\end{tabular}
\end{center}
\end{table}

The parameter $A$ in Table \ref{tbl1} was calculated with the bulk EM. We include corrections to the EM given by Eq. \eqref{eq3}. The formalism presented here is adapted from Ref. \citenum{CostaFilho:2011}. Consider an electron, e, and hole, h, in contracted space ($\gamma$-space) confined by a harmonic potential:
\begin{equation}\label{eq5}
\mathcal{H}=\sum_{i=e,h}\frac{\hat{p}_{\gamma}^2}{2m_{o,i}^*}+\frac{1}{2}m_{o,i}^*\omega_i^2x_i^2;
\end{equation}
where $\omega_i$ is the resonant frequency. The harmonic potential is chosen as a first approximation to discuss the essential features of this model, because the exact form of the confinement potential is not known due to the complicated electronic structure at the interface. We define $\gamma$ to be related to the Gaussian width parameter $\sigma$ according to:
\begin{equation}\label{eq6}
\begin{array}{ll}
&\gamma = \frac{1}{\sigma}\\
\text{with} \;& \sigma ^2 = \frac{\hbar}{m^*_o\omega}.
\end{array}
\end{equation}
The full details of the formalism are in Ref. \citenum{Barbagiovanni:2013_1}. We define the harmonic potential at the interface to solve for $\omega$. Therefore, let $x=D$ in Eq. \eqref{eq5}, where $D$ is the thickness of the QW, and define the confinement energy as the difference between the respective band energies in the QW and the matrix material:
\begin{equation}\label{eq7}
\frac{1}{2}m_{o,j}^*\omega_j^2D^2=\abs{E_{j,SiO_2}-E_{j,Si}(D)}, \; j=CB,VB;
\end{equation}
where, $E$, is the energy in either SiO$_2$ or Si of the e or h in the $CB$ or $VB$, respectively. $E_{CB,SiO_2}$=3.1 eV and $E_{VB,SiO_2}$=4.6 eV with respect to the Si CB minimum or VB maximum, respectively \cite{Seguini:2011}. This model assumes an abrupt interface, which is not always true at the Si/SiO$_2$ interface \cite{Lockwood:1999, deSousa:2002}, however, the essential physics is not lost. Since we are not concerned with the description of a particular theory of QC, we are free to use the simplest case. Take $E_{j,Si}(D)$ as described by Eq. \eqref{eq4}; from Table \ref{tbl1}: $A=0.35$ and $0.66$ for $j=CB$ and $VB$, respectively.

Equations \eqref{eq6}, \eqref{eq7}, and $E_{j,Si}(D)$ express $\gamma$ as a function of the QW thickness, $\gamma(D)$. Equation \eqref{eq1} states that for a larger value of $\gamma$ there is a stronger coupling between the displacement of the carrier particle and the original position. Ref. \citenum{CostaFilho:2011} demonstrates for $\gamma>0$ there is a reduction of the average particle position from the centre of the QW, i.e. $\avg{x}<0.5D$. Therefore, $\gamma$ represents coupling with the interface causing the carrier particles to reside on average closer to the interface than the centre of the QW. Physically, coupling with the interface arises from the polarization effect due to carrier particle self-energy at the NS boundary \cite{Kupchak:2006}. This observation is in agreement with the hypothesis of QC, which implies delocalization in $\v{k}$-space, thus increasing the probability of interface scattering. However, an additional effect is experimentally observed in the case of the hole. As NS dimension is reduced holes become more localized than electrons \cite{Seino:2012}, due to pinning with interface states \cite{Saar:2005, Barbagiovanni:2012}. Therefore, the confinement energy, Eq. \eqref{eq7}, for the hole should not depend on the dimension of the QW, instead we chose \cite{Saar:2005, Garrido:2000}:
\begin{equation}\label{eq8}
E_{VB,Si}=\hbar\omega = 0.13 eV; 
\end{equation}
which models interface coupling with the Si-O-Si vibrational mode (see Sec. \ref{disc}) \cite{Iacona:2000, Martinu:2010, Hadjisavvas:2007}. 

To properly consider Eq. \eqref{eq3} in the context of QC, we express the EM as a function of QW thickness. Since QC implies localization of carrier particles, let $x\rightarrow \avg{x}$ (average position) in Eq. \eqref{eq3}. This replacement ensures that our model considers only the effect of carrier coupling with the interface through the $\gamma$ parameter. For the sake of simplicity, consider an infinite confinement model. The normalized ground state wave-function for infinite confinement in $\gamma$-space is given in Ref. \citenum{CostaFilho:2011}. We find:
\begin{equation}\label{eq9}
\avg{x(D)}=\frac{\gamma D- \ln(1+\gamma D)}{\gamma\ln(1+\gamma D)}- \frac{D\ln(1+\gamma D)}{4\pi^2+ \ln^2(1+\gamma D)}.
\end{equation}

The EMA is well suited to model the Si/SiO$_2$ interface \cite{Gusev:2013} through the use of the envelope function approximation. To test what effect a SPDEM has on QC, it was used as a correction term for Eq. \eqref{eq4}, see Sec. \ref{disc}. We choose a definition for the SPDEM that reflects both the confinement barrier \cite{Khordad:2010, JohnPeter:2008, Quang:2008} and hole pinning:
\begin{equation}\label{eq10}
m_{e}(D)=m_{o,e}^*\left(1-\frac{1}{(1+\gamma_e(D) \avg{x_e(D)})^2}\right)
\end{equation} 
\begin{equation}\label{eq11}
\begin{array}{ll}
&m_{h}(D)=m_{o,h}^*\left(1+\frac{1}{(1+\gamma_h(D) \avg{x_h(D)})^2}\right)\\
\text{or}&m_{h}(D)=m_{o,h}^*\left(1+\frac{1}{(1+\gamma_h \avg{x_h(D)})^2}\right) \; \text{for fixed} \; \gamma_h
\end{array}
\end{equation} 
In these definitions, as $D$ goes to infinity $m(D)$ approaches the bulk EM value. As $D$ goes to zero the electron is increasingly delocalized in \v{k}-space, while the opposite effect holds for the hole, in accordance with experimental observations.

\section{EXPERIMENT}\label{expt}
We cite the results of several experimental works including our own from the University of Western Ontario and from the National Research Council Ottawa, in Sec. \ref{results}. The essential features of each experiment are given here. The details of the experiments can be found in the references provided.

\section{RESULTS}\label{results}

In this section, we present results for Si QDs and QWs, further results can be found in Ref. \citenum{Barbagiovanni:2012}.

\subsection{Quantum Well}\label{QW}

\begin{figure}
\begin{center}
\begin{tabular}{c}
\includegraphics[height=7cm]{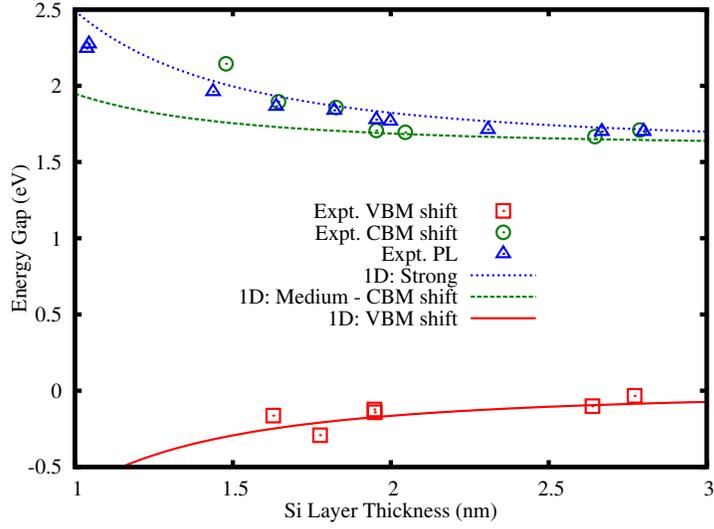} 
\end{tabular}
\end{center}
\caption{Disordered Si-QW data and theoretical fit. Experimental data from Ref. \citenum{Lu:1995}. Theoretical fit using $A$=0.89 and $E_{G}(\infty)=1.6$ eV in Eq. \eqref{eq4}. NB: The CBM shift is offset by the $E_{G}(\infty)$.\label{fig1}}
\end{figure}

Si/SiO$_2$ superlattice Si-QWs have been grown using molecular beam epitaxy, determined to be disordered via Raman scattering measurements, and their thickness found using transmission electron microscopy (TEM) and X-ray diffraction (XRD).\cite{Lu:1995, Lockwood:1996} The change in the valence band maximum (VBM) and conduction band minimum (CBM) position was measured using XPS and Si L$_{2,3}$ edge absorption spectroscopy, respectively, and room temperature photoluminescence (PL) spectroscopy was measured. Fig. \ref{fig1} plots the model predictions with the experimental data. 

In Ref. \citenum{Lu:1995} the authors used a fitting procedure according to the effective mass theory for the $\Delta$E$_{VBM(CBM)}$, resulting in $\Delta$E$_{VBM}=-0.5/D^2$ and $\Delta$E$_{CBM}=0.7/D^2$, where $D$ is the thickness of the QW. Our model predicts $\Delta$E$_{VBM}=-0.66/D^2$ and $\Delta$E$_{CBM}=0.35/D^2$. The trend for $\Delta$E$_{CBM}$ is more accurately given in Ref. \citenum{Lu:1995}. In Ref. \citenum{Lockwood:1996}, the change in $E_G$ was fitted with $A=0.7$ and $E_{G}(\infty)$=1.6 eV, as in Eq. \eqref{eq4}. The fit also determined the effective mass to be m$^*_{h(e)}\approx 1$. The model uses $E_{G}(\infty)$=1.6 eV to fit the experimental PL data well when employing the curve for strong confinement with $A=0.89$. 

Next we look at c-Si/SiO$_2$ QWs fabricated by chemical and thermal processing of silicon-on-insulator wafers.\cite{Lu:2002} The same methods described above were used to determine experimentally the $\Delta$E$_{VBM(CBM)}$ and the change in the gap energy including the total electron yield for a better signal to noise ratio. The thickness of the Si layer was determined by XPS using a mean free path in Si of $\sim$1.6 nm. Note that a thickness of 0.5 nm corresponds to a single unit cell of Si. Therefore, experimental data below $\approx$ 1 nm should be treated with caution. In a parallel study, these c-Si/SiO$_2$ QWs were investigated optically.\cite{Lockwood:2003}
\begin{figure}
\begin{center}
\begin{tabular}{c}
\includegraphics[height=7cm]{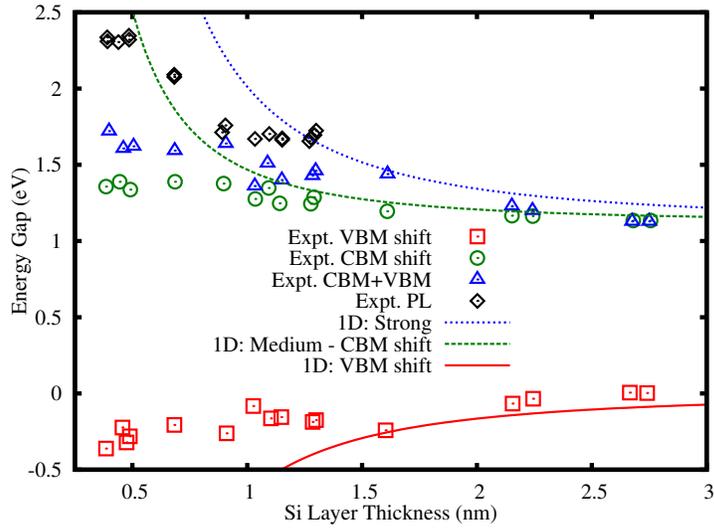} 
\end{tabular}
\end{center}
\caption{Crystalline Si-QW data and theoretical fit. Experimental data from Ref. \citenum{Lu:2002}. Experimental PL data from Ref. \citenum{Lockwood:2003}. Theoretical fit using $A$=0.35 and $E_{G}(\infty)=1.12$ eV in Eq. \eqref{eq4}. NB: The CBM shift is offset by the $E_{G}(\infty)$.\label{fig2}}
\end{figure}

Fig. \ref{fig2} compares experimental measurements and the model results for c-Si-QWs. The $E_G(\infty)$ in the model is 1.12 eV and the $\Delta$E$_{VBM}$ is not significant below 1.5 nm. The $\Delta$E$_{CBM}$, $\Delta$E$_{CBM+VBM}$, and the experimental PL are all well fitted by the curve for medium confinement, with $A=0.35$. In Ref. \citenum{Lockwood:2003} it was found that there is a second PL peak fixed with respect to the Si layer thickness at 1.8 eV. This second peak was associated with interface states. Therefore, we can assign the experimental PL data in Fig. \ref{fig2} with direct e-h recombination modelled by medium confinement.  

\subsection{Quantum Dots}\label{QD}
First we consider  Si QDs formed by ion implantation in SiO$_2$ films, followed by high-temperature annealing in N$_2$ and forming gas.\cite{Mokry:2009} Ref. \citenum{Mokry:2009} reports the QD diameter and crystalline structure observed by TEM, and room temperature PL measurements. TEM data show a Gaussian distribution in the Si-QD diameter with depth, resulting in a stretched exponential PL dynamic.\cite{Mokry:2009,Linnros:1999} 

We compare ion-implanted Si-QDs with Si QDs in a SiO$_2$ matrix prepared by microwave plasma decomposition (MPD) creating ultrafine and densely packed Si QDs\cite{Takagi:1990} (implying that tunnelling effects are important here \cite{Kamenev:2004}). The crystallinity and size was determined by TEM imaging and XRD, respectively. In Ref. \citenum{Takagi:1990}, the authors note that PL was not observed unless the Si QDs were oxidized, implying that surface bonds were passivated with suboxide states eventually forming a surround SiO$_2$ matrix.

\begin{figure}
\begin{center}
\begin{tabular}{c}
\includegraphics[height=7cm]{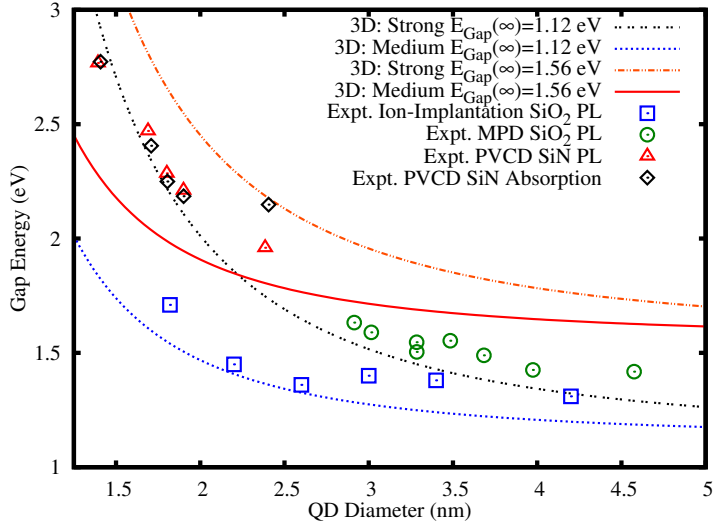} 
\end{tabular}
\end{center}
\caption{Crystalline and amorphous Si-QD data and theoretical fit. `Expt. Ion-Implantion SiO$_2$' refers to crystalline Si QDs embedded in SiO$_2$ from Ref. \citenum{Mokry:2009}. `Expt. microwave plasma decomposition (MPD) SiO$_2$' refers to crystalline Si QDs embedded in SiO$_2$ from Ref. \citenum{Takagi:1990}. `Expt. plasma enhanced chemical vapour deposition (PCVD) SiN' refers to amorphous Si QDs embedded in SiN from Ref. \citenum{Park:2001}. Theoretical fit using $A$=3.57 and 1.39 and $E_{G}(\infty)=1.12\; \text{or}\; 1.56$ eV (as labeled) in Eq. \eqref{eq4}. NB: The absorption data is offset by the $E_{G}(\infty)$.\label{fig3}}
\end{figure}

Fig. \ref{fig3} shows the experimental PL data for ion-implantated and MPD Si QDs together with our calculated curves for strong and medium confinement. Above 3 nm both sets of experimental data follow closely the model of strong confinement with $A$=3.57 and $E_G(\infty)$=1.12 eV. This indicates that for sample diameters larger than this size tunnelling effects are significant, implying a de-localization of carrier states. Iacona et al. measured a similar trend for experimental PL data. \cite{Iacona:2000} Below 3 nm, when QC effects are particularly strong, the ion-implantation data follows the curve for medium confinement, with $A$=1.39. 

Next we consider a-Si QDs embedded in a SiN matrix.\cite{Park:2001}  The Si QDs were fabricated using plasma enhanced chemical vapour deposition. The size and amorphous structure were measured using TEM and the PL was taken at room temperature. Absorption data was taken by ultraviolet-visible absorption spectroscopy. The value for the bulk band gap given by the authors is 1.56 eV, which is obtained via a fitting procedure. This value is known to vary between 1.5$\rightarrow$1.6 eV, for Si samples prepared similarly.\cite{Park:2001} 

We can see in Fig. \ref{fig3} that the experimental data for absorption and PL of a-Si QDs embedded in SiN lies between the curve for medium ($A$=1.39) and strong ($A$=3.57) confinement, with $E_{G}(\infty)$=1.56. Using a fitting procedure, the authors of Ref. \citenum{Park:2001} found $A$=2.40. The authors further conclude that by observing the fact that the experimental absorption data lies close to the PL data, one can conclude that the PL data for these samples is a good measure of the actual change in the $E_G(D)$.\cite{Park:2001} Notice that this situation is similar to that observed for Si-QWs (see Fig. \ref{fig1} and \ref{fig2}).

\section{DISCUSSION}\label{disc}

To summarize the comparisons made in Sec. \ref{results}, we first consider the relationship between experimental absorption and PL data. In the case of disordered-Si-QWs (Fig. \ref{fig1}), c-Si-QWs (Fig. \ref{fig2}) and a-Si-QDs in SiN (Fig. \ref{fig3}) the absorption curve follows closely with the PL. As mentioned in Sec. \ref{QD}, this result indicates that the PL  measurement is an accurate measure of $E_G(D)$. Furthermore, in the case of Si-QWs the VBM does not change significantly. Therefore, we conclude that the model dependence between these three systems does not lie in the change in the VBM. 

PL data for Porous Si QDs (por-Si-QDs)\cite{Lockwood:1995, Lockwood:1994, Barbagiovanni:2012} (not shown here) are nearly identical to the MPD Si-QDs (Fig. \ref{fig3}), which indicates that these systems are structurally similar with similar decay dynamics. In the case of por-Si it has been found that this system is under tensile stress.\cite{Kun:2005} Tensile stress, which is a function of the thickness of oxide, is known to increase the band gap.\cite{Hong:2003} It is known that the surrounding oxide has a strong effect on the resulting PL in por-Si.\cite{Wolkin:1999} The resulting Si-O-Si bonds due to the oxidation process place large stresses on the por-Si crystallites. In addition, it has been shown that the dominant PL comes from surface states.\cite{Lockwood:1994} At the surface or interface states, it has been shown that band bending on the order of 0.2$\rightarrow$0.3 eV can occur.\cite{Svechnikov:1998}
 
Ion implanted Si-QDs (Fig. \ref{fig3}) and Ge-QDs\cite{Takeoka:1998, Barbagiovanni:2012} (not shown here) have the same behaviour above 3 nm. They lie close to the curve for strong confinement, similar to the case of por-Si, indicating that possible stresses or interface states are important in this regime. Ge is known to experience stress in a SiO$_2$ matrix.\cite{Sharp:2005} Tensile stress can be relieved depending on the nature of the interface bonds and the surface to volume ratio of Si:SiO$_2$.\cite{Hong:2003} In the work of Ref. \citenum{Barbagiovanni:2011} it was found from Raman spectroscopy that ion-implanted QDs are not under stress for diameters smaller than 3 nm. Therefore, c-Si-QDs produced by ion implantation and c-Ge-QDs are well modelled by medium confinement below 3 nm.

Finally, a-Si-QDs in SiN (Fig. \ref{fig3}) lie between medium and strong confinement (see Sec. \ref{QD}). SiN has a band gap of 5.3 eV versus SiO$_2$ at 9.2 eV, which allows for tunnelling of carrier states.\cite{Park:2001} More importantly, if we consider the nucleation process during thermal annealing and consider the bond enthalpies for diatomic species (SiN at 470 kJ/mol and SiO at 799 kJ/mol), it is easier to break SiN bonds, thus allowing for a greater degree of intermixing at the QD-matrix interface. Therefore, a SiN matrix acts more like a finite potential barrier, which lowers the gap energy from the infinite case. A numerical computation indicates that the difference between the case of finite versus infinite confinement potential is between 10$\%$ and 15$\%$ depending on the size of the system. This difference exactly corresponds with the difference we see in Fig. \ref{fig3}. Therefore, we conclude that a-Si-QDs in SiN are well modelled by strong confinement. 

From the results above and considering modifications that must be made to our model to account for non-direct e-h recombination phenomena, it is clear that strong confinement describes a-materials and medium confinement describes c-materials. Therefore, since QC of a particle is a function of the delocalization of that particle with respect to the dimension of the system, we need to account for the fact that the hole becomes more delocalized in the a-system than in the c-system. This fact may or may not be seen as a shift in the VBM. As noted above, disordered-Si-QWs, c-Si-QWs and a-Si-QDs in SiN all do not show a large variation in the VBM. 

A mechanism for pinning of the hole states in c-Si-QDs was discussed in the work of Sa'ar et al. as a function of the hole coupling with vibrons.\cite{Saar:2005} However, this phenomenon does not account for the fact that the hole becomes more delocalized in the a-system. It is well known that band-tail states play a very important role in the band structure of a-materials, even though the population density is relatively low.\cite{Street:1991} Kanemitsu et al. (and Refs. within), report the experimental observation that the band-tail states become strongly delocalized in the a-system, while the hole remains relatively localized in the c-system.\cite{Kanemitsu:2002} This observation accounts for what is observed in this work. The effect of hole pinning is modelled in the SPDEM using Eq. \eqref{eq8}.

\begin{figure}
\begin{center}
\begin{tabular}{c}
\includegraphics[height=7cm]{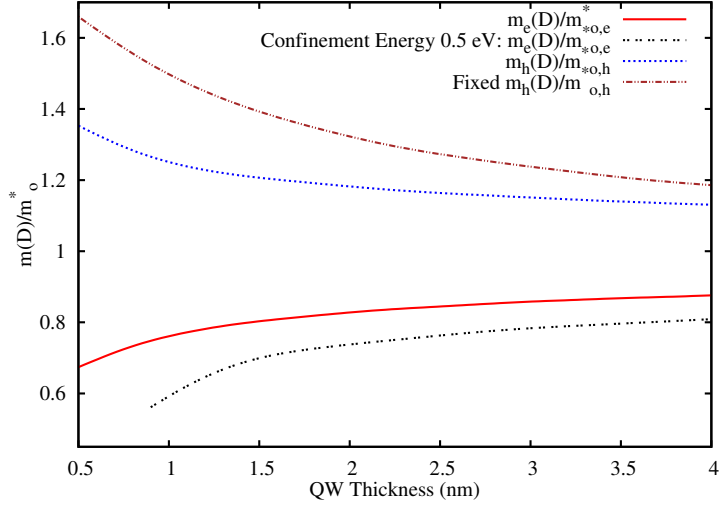}
\end{tabular}
\end{center}
\caption{Variation of $m_{e}(D)$ and $m_{h}(D)$ as a function of QW thickness with respect to the bulk EM. `Confinement energy 0.5 eV' for $m_{e}(D)$ is for a lowered potential barrier, in accordance with a Si/Ge QW. $m_{h}(D)$ is plotted for a variable and fixed confinement energy.\label{fig4}}
\end{figure}

Next, we consider how variations in the EM effect the confinement strength. The results of Eqs. \eqref{eq10} and \eqref{eq11} are shown in Fig. \ref{fig4} with respect to the bulk effective mass. The electron SPDEM is reduced from the bulk value and decreases as the QW thickness is reduced. This effect increases the electron energy and the tunnelling probability. The opposite effect is seen for the hole. There are no experimental reports for the hole EM in a QW structure, but the electron EM does show the same experimental trend as we find here \cite{Lockwood:1996}. In the work of Ref. \citenum{Rossner:2003}, the authors report a reduction of the electron EM from the bulk value (1.08$m_o$ and 0.56$m_o$ in Si and Ge, respectively) at 0.08$m_o$ ($m_o$ is the free electron mass) by fitting temperature dependent Shubnikov-de Haas oscillations for Ge QWs. The tunnelling EM is reported to be 0.09$m_o$ in amorphous Si \cite{Shannon:1993}. Temperature dependent PL measurements place the electron EM at 0.014$m_o$ for Ge/Si superlattices \cite{Yang:2004}.
 
This issue of the correct effective mass is more poignant when considering the $\Delta$E$_{CBM(VBM)}$. In this work (Table \ref{tbl1}), $\Delta$E$_{VBM}>\Delta$E$_{CBM}$, which is understood, because the effective mass of the hole is smaller than the electron. However, experiment consistently shows the opposite effect, see Figs. \ref{fig1},\ref{fig2} and see Refs. \citenum{Afanasev:2009, Seguini:2011}. This observation implies that experiment is measuring a larger decrease in the electron effective mass than the hole, or possibly a relative increase in the hole mass compared to the electron. This observation is nearly consistent with Ref. \citenum{Seino:2011}, where they predict a nearly symmetric change. Furthermore, recall that experiment reports a decrease in the electron effective mass.\cite{Rossner:2003,Yang:2004}  
 
The decrease in the electron EM and increase in the hole EM is consistent for the crystalline system with our observation of medium confinement, because the hole is more spatially localized. In being consistent with experiment, we dropped the hole contribution for the crystalline system in the ideal approximation, because this term is not as significant as the electron according to the $\Delta$E$_{CBM(VBM)}$ measurements, described above. Although, there may still be a slight hole contribution in this ideal approximation, which needs further study. In addition, in our theoretical modelling, we have consistent results for strong confinement in the amorphous samples, because both the electron and hole effective mass decrease implying confinement of both, due to spatial de-localization\cite{Singh:2002, Barbagiovanni:2012}. Although, the relative contribution from the electron versus the hole is not clear and needs further study.
 
Experiment reports a larger reduction of the effective mass than we see in our current model. The reason for the discrepancy is because experimental Ge/Si QWs have a lower confinement energy than what is given in Eq. \eqref{eq7}. Replacing $E_{CB,SiO_2}\rightarrow E_{CB,Si}$ and $E_{CB,Si}\rightarrow E_{CB,Ge}$ to model a Ge QW confined by Si spacers gives a lowered initial confinement energy of 0.5 eV and further reduces the electron EM, see Fig. \ref{fig4}. Therefore, our model does capture the correct features of the confinement barrier. A similar trend is seen in the work of Ref. \citenum{Lepadatu:2010} for the electron EM as a function of the potential barrier. Calculations that strictly consider the variation of the EM with QW thickness, i.e. no spatial dependence, find a decrease in the electron EM and an increase in the hole EM with dimension and always with a value larger than the bulk EM \cite{Seino:2011}. On the the other hand, the SPDEM model yields a value for the electron EM reduced from the bulk EM value. Furthermore, this model qualitatively agrees with the description of tunnelling in Ref. \citenum{Seino:2012}.

\begin{figure}
\begin{center}
\begin{tabular}{c}
\includegraphics[height=7cm]{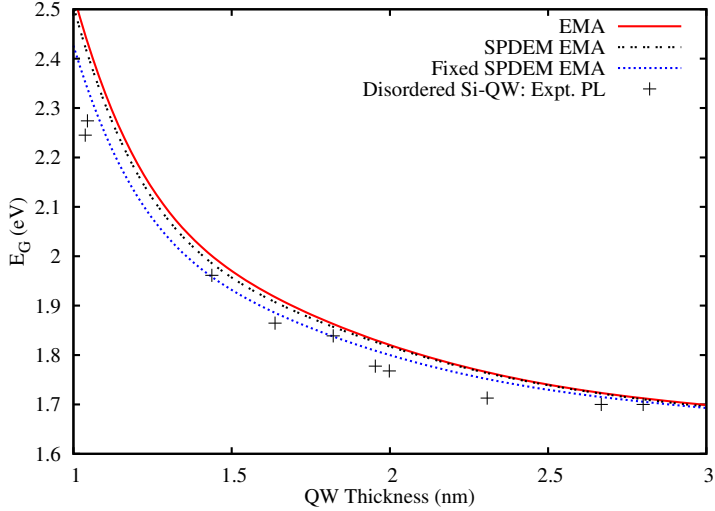}
\end{tabular}
\end{center}
\caption{Variation of $E_G$ as a function of QW thickness. `EMA' is given by Eq. \eqref{eq4} with $A$=0.89. `SPDEM EMA' is given by Eq. \eqref{eq4} and corrected with Eqs. \eqref{eq10} and \eqref{eq11}, including the results for a fixed $\gamma_h$. `Disordered Si-QW: Expt. PL' is from Ref. \citenum{Lu:1995} and was obtained from PL measurements.
\label{fig5}}
\end{figure}

Eqs. \eqref{eq10} and \eqref{eq11} are used to correct $E_G(D)$ given by Eq. \eqref{eq4} for a-Si QWs with $A$=0.89. We compare this model shown in Fig. \ref{fig5} with the experimental data of Ref. \citenum{Lu:1995} from Fig. \ref{fig1}. In Fig. \ref{fig5}, Eq. \ref{eq4} is labelled `EMA' and labelled `SPDEM EMA' when corrected for by \eqref{eq10} and \eqref{eq11}, including the results for a fixed $\gamma_h$. Under the current model, there is not a significant change between the SPDEM EMA and EMA results. This result is because the electron SPDEM increase the $E_G$ and the hole SPDEM decreases the $E_G$. In the case of fixed hole energy, there is a slightly larger decrease in the $E_G$. It is clear that the effect of a SPDEM in our QC model increases the accuracy with experiment. Therefore, we conclude that a SPDEM acts to reduce the confinement energy due to coupling with the interface, which simulates the effect of a reduced confinement barrier. Furthermore, it has been demonstrated\cite{Gusev:2013} that the effect carrier tunnelling into the oxide matrix is small, in agreement with the results in Fig. \ref{fig5}.

\section{CONCLUSIONS}\label{conl}
We have studied the effect of confinement dimensions and crystallinity on the magnitude of the band gap expansion (as a function of decreasing size) in Si and Ge NSs (quantum wells (QWs), wires (Q-wires) and dots (QDs). Medium and strong confinement models provide the best fit to experimental results; moreover crystalline materials exhibit medium confinement, while amorphous materials exhibit strong confinement regardless of the confinement dimensions of the system. This difference in confinement strength was explained by considering the extent of spatial delocalization of the hole. A possible explanation is hole pinning due to coupling with the vibronic states.\cite{Saar:2005} It has previously been reported\cite{Kanemitsu:2002} that band tail states become strongly delocalized in the amorphous system compared to the crystalline system. This hole delocalization would partially account for the trends observed in our work. A lower value of the effective mass is reported for the amorphous system, which accounts for the trends observed in our work, while the hole mass increases and the electron mass decreases as a function of spatial confinement.\cite{Seino:2011, Rossner:2003, Yang:2004} We have studied the effect of a SPDEM with one singularity in a two band EM model for QC by introducing a phenomenological model with an inverse characteristic length related to the Gaussian width parameter. This model for the EM accounts for electron coupling with the interface and allows for tunnelling effects to be included in the confinement energy. As the thickness of a QW is reduced, there is increased electron coupling with the interface, which increases the tunnelling probability. Furthermore, we find these results to be a promising first step toward the development of better models of the effective mass parameter in low-dimensional theories. More work needs to be done to test the results of Eq. \eqref{eq2} within more sophisticated theories, such as, $\v{k}\cdot\v{p}$, tight-binding, or pseudo-potential methods. 


\begin{thebibliography}{10}

\bibitem{edit:2010}
editorial {\em Nat. Nanotechnol.}~{\bf 5}, p.~381, 2010.

\bibitem{Lockwood:2004}
D.~J. Lockwood and L.~Pavesi, {\em Silicon Photonics}, vol.~94, pp.~1--52.
\newblock Springer, Berlin, 2004.

\bibitem{Loss:1998}
D.~Loss and D.~P. DiVincenzo {\em Phys. Rev. A.}~{\bf 57}, p.~120, 1998.

\bibitem{Yoffe:2002}
A.~D. Yoffe {\em Adv. Phys.}~{\bf 51}, p.~799, 2002.

\bibitem{Frensley:1977}
W.~R. Frensley and H.~Kroemer {\em Phys. Rev. B.}~{\bf 16}, p.~2642, 1977.

\bibitem{Bir:1974}
G.~L. Bir and G.~E. Pikus, {\em Symmetry and Strain-Induced Effects in
  Semiconductors}, Wiley, New York, 1976.

\bibitem{Valentin:2007}
A.~Valentin, J.~S\'{e}e, S.~Galdin-Retailleau, and P.~Dollfus {\em J. Phys.:
  Conf. Ser.}~{\bf 92}, p.~012048, 2007.

\bibitem{VanDao:2005}
L.~V. Dao, X.~Wen, M.~T.~T. Do, P.~Hannaford, E.~C. Cho, Y.~H. Cho, and
  Y.~Huang {\em J. Appl. Phys.}~{\bf 97}, p.~013501, 2005.

\bibitem{Heiss:2005}
W.~D. Heiss, {\em Quantum Dots: a Doorway to Nanoscale Physics}, Springer,
  Berlin, 2005.

\bibitem{Zunger:2001}
A.~Zunger {\em Phys. Stat. Sol. (b)}~{\bf 224}, p.~727, 2001.

\bibitem{Tran:1990}
D.~B.~T. Thoai, Y.~Z. Hu, and S.~W. Koch {\em Phys. Rev. B.}~{\bf 42},
  p.~11261, 1990.

\bibitem{Tit:2010}
N.~Tit, Z.~H. Yamani, J.~Graham, and A.~Ayesh {\em Mater. Chem. Phys.}~{\bf
  24}, p.~927, 2010.

\bibitem{Buuren:1998}
T.~van Buuren, L.~N. Dinh, L.~L. Chase, W.~J. Siekhaus, and L.~J. Terminello
  {\em Phys. Rev. Lett.}~{\bf 80}, p.~3803, 1998.

\bibitem{Cosentino:2013}
S.~Cosentino, M.~Miritello, I.~Crupi, G.~Nicotra, F.~Simone, C.~Spinella,
  A.~Terrasi, and S.~Mirabella {\em Nanoscale Res. Lett.}~{\bf 8}, p.~128,
  2013.

\bibitem{vonRoos:1983}
O.~von Roos {\em Phys. Rev. B.}~{\bf 27}, p.~7547, 1983.

\bibitem{Young:1989}
K.~Young {\em Phys. Rev. B.}~{\bf 39}, p.~13434, 1989.

\bibitem{Geller:1993}
M.~R. Geller and W.~Kohn {\em Phys. Rev. Lett.}~{\bf 70}, p.~3103, 1993.

\bibitem{Bastard:1981}
G.~Bastard {\em Phys. Rev. B.}~{\bf 24}, p.~5693, 1981.

\bibitem{Ganguly:2006}
A.~Ganguly, {\c{S}.}.~Kuru, J.~Negro, and L.~M. Nieto {\em Phys. Lett. A}~{\bf
  360}, p.~228, 2006.

\bibitem{Borovitskaya:2000}
E.~Borovitskaya and M.~S. Shur {\em Solid-State Electron.}~{\bf 44}, p.~1609,
  2000.

\bibitem{Moskalenko:2007}
A.~S. Moskalenko, J.~Berakdar, A.~A. Prokofiev, and I.~N. Yassievich {\em Phys.
  Rev. B.}~{\bf 76}, p.~085427, 2007.

\bibitem{Tomic:2011}
S.~Tomi\'{c} and N.~Vukmirovi\'{c} {\em J. Appl. Phys.}~{\bf 110}, p.~053710,
  2011.

\bibitem{Delerue:2004}
C.~Delerue and M.~Lannoo, {\em Nanostructures: Theory and Modelling}, Springer,
  Berlin, 2004.

\bibitem{Niquet:2000_1}
Y.~M. Niquet, C.~Delerue, G.~Allan, and M.~Lannoo {\em Phys. Rev. B.}~{\bf 62},
  p.~5109, 2000.

\bibitem{Seino:2012}
K.~Seino, F.~Bechstedt, and P.~Kroll {\em Phys. Rev. B.}~{\bf 86}, p.~075312,
  2012.

\bibitem{Barbagiovanni:2012}
E.~G. Barbagiovanni, D.~J. Lockwood, P.~J. Simpson, and L.~V. Goncharova {\em
  J. Appl. Phys.}~{\bf 111}, p.~034307, 2012.

\bibitem{Seas:1997}
A.~Seas and C.~Christoﬁdes {\em Semicond. and Semimetals}~{\bf 46}, p.~39,
  1997.

\bibitem{Seino:2011}
K.~Seino and F.~Bechstedt {\em Semicond. Sci. Technol.}~{\bf 26}, p.~014024,
  2011.

\bibitem{Lockwood:1996}
D.~J. Lockwood, Z.~H. Lu, and J.~M. Baribeau {\em Phys. Rev. Lett.}~{\bf 76},
  p.~539, 1996.

\bibitem{Rossner:2003}
B.~R\"{o}$\ss$ner, G.~Isella, and H.~von K\"{a}nel {\em Appl. Phys. Lett.}~{\bf
  82}, p.~754, 2003.

\bibitem{Barbagiovanni:2013}
E.~G. Barbagiovanni, D.~J. Lockwood, P.~J. Simpson, and L.~V. Goncharova,
  ``{Quantum Confinement in Si and Ge Nanostructures: Theory and Experiment}.''
  to be published Appl. Phys. Rev. (2014).

\bibitem{CostaFilho:2011}
R.~N. {Costa Filho}, M.~P. Almeida, G.~A. Farias, and J.~S. {Andrade Jr.} {\em
  Phys. Rev. A.}~{\bf 84}, p.~050102, 2011.

\bibitem{CostaFilho:2013}
R.~N. {Costa Filho}, G.~Alencar, B.~S. Skagerstam, and J.~S. {Andrade Jr.} {\em
  Europhys. Lett.}~{\bf 101}, p.~10009, 2013.

\bibitem{Barbagiovanni:2011}
E.~G. Barbagiovanni, L.~V. Goncharova, and P.~J. Simpson {\em Phys. Rev.
  B.}~{\bf 83}, p.~035112, 2011.

\bibitem{Barbagiovanni:2013_1}
E.~G. Barbagiovanni and R.~N.~C. Filho, ``{Quantum Confinement in Nonadditive
  Space with a Spatially Dependent Effective Mass for Si and Ge Quantum
  Wells}.'' {arxiv:1311.5335}.

\bibitem{Seguini:2011}
G.~Seguini, S.~Schamm-Chardon, P.~Pellegrino, and M.~Perego {\em Appl. Phys.
  Lett.}~{\bf 99}, p.~082107, 2011.

\bibitem{Lockwood:1999}
D.~J. Lockwood {\em Phase Transitions}~{\bf 68}, p.~151, 1999.

\bibitem{deSousa:2002}
J.~S. de~Sousa, E.~W.~S. Caetanoa, J.~R. Gon\c{c}alves, G.~A. Farias, V.~N.
  Freire, and E.~F. {da Silva Jr.} {\em Appl. Surf. Sci.}~{\bf 190}, p.~166,
  2002.

\bibitem{Kupchak:2006}
I.~M. Kupchak, D.~V. Korbutyak, Y.~V. Kryuchenko, A.~V. Sachenko, I.~O.
  Sokolovski\v{i}, and O.~M. Sreseli {\em Semicond.}~{\bf 40}, p.~94, 2006.

\bibitem{Saar:2005}
A.~Sa'ar, Y.~Reichman, M.~Dovrat, D.~Krapf, J.~Jedrzejewski, and I.~Balberg
  {\em Nano Lett.}~{\bf 5}, p.~2443, 2005.

\bibitem{Garrido:2000}
B.~Garrido, M.~L\'{o}pez, O.~Gonz\'{a}lez, A.~P\'{e}rez-Rodr\'{i}guez, J.~R.
  Morante, and C.~Bonafos {\em Appl. Phys. Lett.}~{\bf 77}, p.~3143, 2000.

\bibitem{Iacona:2000}
F.~Iacona, G.~Franzo, and C.~Spinella {\em J. Appl. Phys.}~{\bf 87}, p.~1295,
  2000.

\bibitem{Martinu:2010}
L.~Martinu, O.~Zabeida, and J.~E. Klemberg-Sapieha, {\em Plasma-Enhanced
  Chemical Vapor Deposition of Functional Coatings}, pp.~392--465.
\newblock Elsevier Inc., Oxford, 3~ed., 2010.

\bibitem{Hadjisavvas:2007}
G.~Hadjisavvas and P.~C. Kelires {\em Physica E}~{\bf 38}, p.~99, 2007.

\bibitem{Gusev:2013}
O.~B. Gusev, A.~N. Poddubny, A.~A. Prokofiev, and I.~N. Yassievich {\em
  Semiconductors}~{\bf 47}, p.~183, 2013.

\bibitem{Khordad:2010}
R.~Khordad {\em Physica E}~{\bf 42}, p.~1503, 2010.

\bibitem{JohnPeter:2008}
A.~{John Peter} and K.~Navaneethakrishnan {\em Physica E}~{\bf 40}, p.~2747,
  2008.

\bibitem{Quang:2008}
N.~H. Quang, N.~T. Truc, and Y.~M. Niquet {\em Comput. Mater. Sci.}~{\bf 44},
  p.~21, 2008.

\bibitem{Lu:1995}
Z.~H. Lu, D.~J. Lockwood, and J.~M. Baribeau {\em Nature}~{\bf 378}, p.~258,
  1995.

\bibitem{Lu:2002}
Z.~H. Lu and D.~Grozea {\em Appl. Phys. Lett.}~{\bf 80}, p.~255, 2002.

\bibitem{Lockwood:2003}
D.~J. Lockwood, M.~W.~C. Dharma-wardana, Z.~H. Lu, D.~H. Grozea, P.~Carrier,
  and L.~J. Lewis {\em Mater. Res. Soc. Symp. Proc.}~{\bf 737}, p.~F1.1.1,
  2003.

\bibitem{Mokry:2009}
C.~R. Mokry, P.~J. Simpson, and A.~P. Knights {\em J. Appl. Phys.}~{\bf 105},
  p.~114301, 2009.

\bibitem{Linnros:1999}
J.~Linnros, N.~Lalic, A.~Galeckas, and V.~Grivickas {\em J. Appl. Phys.}~{\bf
  86}, p.~6128, 1999.

\bibitem{Takagi:1990}
H.~Takagi, H.~Ogawa, Y.~Yamazaki, A.~Ishizaki, and T.~Nakagiri {\em Appl. Phys.
  Lett.}~{\bf 56}, p.~2379, 1990.

\bibitem{Kamenev:2004}
B.~V. Kamenev, G.~F. Grom, D.~J. Lockwood, J.~P. McCafrey, B.~Laikhtman, and
  L.~Tsybeskov {\em Phys. Rev. B.}~{\bf 69}, p.~235306, 2004.

\bibitem{Park:2001}
N.~M. Park, C.~J. Choi, T.~Y. Seong, and S.~J. Park {\em Phys. Rev. Lett.}~{\bf
  86}, p.~1355, 2001.

\bibitem{Lockwood:1995}
D.~J. Lockwood and A.~G. Wang {\em Solid State Commun.}~{\bf 94}, p.~905, 1995.

\bibitem{Lockwood:1994}
D.~J. Lockwood {\em Solid State Commun.}~{\bf 92}, p.~101, 1994.

\bibitem{Kun:2005}
L.~Z. Kun, K.~Y. Lan, C.~Hao, H.~Ming, and Q.~Yu {\em Chin. Phys. Lett.}~{\bf
  22}, p.~984, 2005.

\bibitem{Hong:2003}
C.~C. Hong, W.~J. Liao, and J.~G. Hwu {\em Appl. Phys. Lett.}~{\bf 82},
  p.~3916, 2003.

\bibitem{Wolkin:1999}
M.~V. Wolkin, J.~Jorne, P.~M. Fauchet, G.~Allan, and C.~Delerue {\em Phys. Rev.
  Lett.}~{\bf 82}, p.~197, 1999.

\bibitem{Svechnikov:1998}
S.~V. Svechnikov, E.~B. Kaganovich, and E.~G. Manoilov {\em Semi. Phy. Quantum
  Elect. Optoelect.}~{\bf 1}, p.~13, 1998.

\bibitem{Takeoka:1998}
S.~Takeoka, M.~Fujii, S.~Hayashi, and K.~Yamamoto {\em Phys. Rev. B.}~{\bf 58},
  p.~7921, 1998.

\bibitem{Sharp:2005}
I.~D. Sharp, D.~O. Yi, Q.~Xu, C.~Y. Liao, J.~W. Beeman, Z.~Liliental-Weber,
  K.~M. Yu, D.~N. Zakharov, J.~W. Ager, D.~C. Chrzan, and E.~E. Haller {\em
  Appl. Phys. Lett.}~{\bf 86}, p.~063107, 2005.

\bibitem{Street:1991}
R.~A. Street, {\em Hydrogenated Amorphous Silicon}, Cambridge University Press,
  Cambridge, 1991.

\bibitem{Kanemitsu:2002}
Y.~Kanemitsu {\em J. Luminescence}~{\bf 100}, p.~209, 2002.

\bibitem{Shannon:1993}
J.~M. Shannon and K.~J. B.~M. Nieuwesteeg {\em Appl. Phys. Lett.}~{\bf 62},
  p.~1815, 1993.

\bibitem{Yang:2004}
Z.~Yang, Y.~Shi, J.~Liu, B.~Yan, R.~Zhang, Y.~Zheng, and K.~Wang {\em Mater.
  Lett.}~{\bf 58}, p.~3765, 2004.

\bibitem{Afanasev:2009}
V.~V. Afanas’ev, M.~Badylevich, A.~Stesmans, A.~Laha, H.~J. Osten, and
  A.~Fissel {\em Appl. Phys. Lett.}~{\bf 95}, p.~102107, 2009.

\bibitem{Singh:2002}
J.~Singh {\em J. Non-Cryst. Solids}~{\bf 299}, p.~444, 2002.

\bibitem{Lepadatu:2010}
A.~M. Lepadatu, I.~Stavarache, M.~L. Ciurea, and V.~Iancu {\em J. Appl.
  Phys.}~{\bf 107}, p.~033721, 2010.

\end{thebibliography}
\bibliographystyle{spiebib}

\end{document}